\documentstyle[12pt]{article}
\newcommand{\be}{\begin{equation}}
\newcommand{\ee}{\end{equation}}
\newcommand{\ba}{\begin{eqnarray}}
\newcommand{\ea}{\end{eqnarray}}

\begin{document}
\begin{center}
{\bf SUSY QUANTUM MECHANICS WITH COMPLEX SUPERPOTENTIALS AND
REAL ENERGY SPECTRA }\\
\vspace{1cm}
A. A. ANDRIANOV\footnote{{\it Department of Theoretical
Physics, University of Sankt-Petersburg,198904 Sankt-Petersburg, Russia.
E-mail:} andrian@snoopy.phys.spbu.ru;\hspace{3ex}
ioffe@snoopy.phys.spbu.ru},
F. CANNATA\footnote{{\it Dipartimento di Fisica and INFN, Via Irnerio 46,
40126 Bologna, Italy. E-mail:} cannata@bo.infn.it},
J.-P. DEDONDER\footnote{{\it GMPIB, Universit\'e Paris7 - Denis Diderot,
Case 7021, 2 place Jussieu, F75251 Paris Cedex 05 and Division de Physique 
Th\'eorique, IPN, F91406 Orsay, France. E-mail:} dedonder@paris7.jussieu.fr}
  M. V. IOFFE$^1$, 
\end{center}
\vspace{1cm}
\hspace*{0.5in} \begin{minipage}{5.0in}
{\small 
We extend the standard intertwining relations used in Supersymmetrical 
(SUSY) Quantum
Mechanics which involve real superpotentials to complex superpotentials.
This allows to deal with a large class of non-hermitean Hamiltonians and to
study in general the isospectrality between complex potentials.
In very specific cases we can
construct in a natural way "quasi-complex" potentials
which we define as complex potentials having a global property such as
to lead to a Hamiltonian
with real spectrum. We also obtained a class of complex transparent 
potentials whose Hamiltonian can be intertwined to a free Hamiltonian. 
 We provide a variety of examples 
both for the radial problem (half axis)
and for the standard one-dimensional problem (the whole axis), including
remarks concerning scattering problems. 
} 

\end{minipage}
\section*{\large\bf 1.\quad Introduction and general formalism}
\hspace*{2ex} 

Non-hermitean interactions have been discussed in Field Theory and
Statistical Mechanics, see for instance \cite{statistical}, with 
applications to Condensed Matter \cite{nonsymmetrical} and to 
Biology \cite{biology}. In
the context of Hadronic and Nuclear Physics, the so-called optical
potential framework \cite{optical} also involves this type of Hamiltonians
as studied by Baye et al. \cite{baye}: the imaginary part in the potential
signals the opening of new channels. In the present paper we investigate
this class of systems within the Supersymmetrical (SUSY) 
approach in Quantum Mechanics with particular attention to complex
potentials associated to a real spectrum.
Recently there was a lively activity on complex potentials with 
the property of PT (parity and time reversal) invariance, associated to
real and positive spectra \cite{new}. 

In the standard approach to SUSY Quantum Mechanics one makes use of
intertwining relations where the intertwined Hamiltonians are
hermitean and the intertwining operators depend on real "superpotentials"
\cite{review}.

Here we generalize this approach to complex superpotentials $W(y)$
which allow to deal with non-hermitean Hamiltonians.
Our approach makes use of supercharges with complex superpotentials
intertwining two partner Hamiltonians \cite{acdi}:  
\be
H^{(0)} q^+ = q^+ H^{(1)} . \label{intertw} 
\ee
The Hamiltonians $H^{(0)}, H^{(1)}$ contain complex potentials 
$V^{(0)}(y) \equiv U^{(0)}_R(y) + i U^{(0)}_I(y)$ and 
$V^{(1)}(y) \equiv U^{(1)}_R(y) + i U^{(1)}_I(y)$
and the supercharge $q^+$ is defined by
\be
q^+ = -\partial + W(y) = - \partial + f(y) + i g(y) \label{supercharge}
\ee
with  $f(y) ,g(y)\,$ real functions. 
The role of the intertwining operator $q^+$ is to map solutions of the 
Schroedinger equation corresponding to $H^{(1)}$ to solutions 
corresponding to $H^{(0)}$ for identical spectral parameters
$\Psi^{(0)}(y) \sim q^+ \Psi^{(1)}(y)$.

By comparing (\ref{intertw}) with the factorization approach 
\cite{review}, \cite{baye}
one can explicitly derive that (\ref{intertw}) is equivalent to 
the factorization at a complex energy 
$E = \epsilon_R + i\epsilon_I $ of the above mentioned Hamiltonians 
\footnote{We know however that 
intertwining is more general than factorization as one can check
from matrix SUSY Quantum Mechanics \cite{acin} and from Higher order SUSY
Quantum Mechanics \cite{acdi}.}. 
Furthermore by taking successively the hermitean and complex conjugations 
a second intertwining relation can be derived:
\be
q^-H^{(0)} = H^{(1)}q^- . \label{intertw2} 
\ee
The supercharge 
\be
q^- = \partial + W(y) = \partial + f(y) + i g(y)  \label{supercharge2}
\ee
is related to $q^+$ by successive hermitean and complex 
conjugations.
It maps solutions of the 
Schroedinger equation corresponding to $H^{(0)}$ to solutions 
corresponding to $H^{(1)}$ for identical spectral parameters
$\Psi^{(1)}(y) \sim q^- \Psi^{(0)}(y)$.

In general one expects that these operators $ q^{\pm} $ will not map physical
wave functions into physical ones. We will study the conditions
allowing to realize such a mapping apart from one state, solution of
the equation $ q^{\pm} \Psi = 0\, , $ called zero mode : from now on we will
refer to such a mapping as to the case of  strict isospectrality.

In \cite{baye} the standard factorization approach to SUSY Quantum
Mechanics was suitably extended to the case of
complex potentials. In the construction of isospectral 
Hamiltonians our emphasis will be more on the role of the
intertwining relations (\ref{intertw}) with complex superpotentials 
$W(y) = f(y)+ig(y)$, including 
the case of intertwining hermitean and non-hermitean Hamiltonians.

Eqs.(\ref{intertw}) and (\ref{intertw2})
yield the following relations between the potentials $V^{(0)}, V^{(1)}$
and the superpotential $W(y) :$
\ba
U^{(0)}_R(y) &=& - f'(y) + f^2(y) - g^2(y) + \epsilon_R; \label{eq1}\\
U^{(0)}_I(y) &=& - g'(y) + 2 f(y) g(y) + \epsilon_I;  \label{eq2}\\
U^{(1)}_R(y) &=& f'(y) + f^2(y) - g^2(y) + \epsilon_R; \label{eq3}\\
U^{(1)}_I(y) &=& g'(y) + 2 f(y) g(y) + \epsilon_I , \label{eq4}
\ea
where $\epsilon_{R,I}$ are real numbers which in the factorization 
approach correspond to the factorization energy 
$E = \epsilon_R + i\epsilon_I$.

These relations express both complex potentials in terms of two arbitrary
real functions $f(y), g(y)$ determining the complex superpotential
\footnote{Indeed $V^{(1)}(y) - V^{(0)}(y) =
[ q^-, q^+ ] = 2W'(y).$ }.
As far as their solution is concerned we would like to stress that in general
it is not possible to consider one potential to be given and the other
to be constructed since this problem in general cannot be solved analytically.
What one can do is to provide interesting
classes of model systems with $U^{(i)}_R, U^{(i)}_I, f, g$ to be 
determined contextually.

The problem can be formulated both as one dimensional (the whole axis, $y 
\equiv x$)
for the wave function $\Psi (x)$
or as a radial problem (half-axis, $y \equiv r$) 
for $r \cdot \psi (r) \equiv \Psi(r)$. 
The conditions for mapping physical
wave functions into physical wave functions can be different for these two
cases.

There is a possibility of imposing $U^{(1)}_I(y) = U^{(0)}_I(y)$ 
or $U^{(1)}_R(y) = U^{(0)}_R(y)$ or more complicated conditions
since we deal with two functions $f,g$.
The first conditions leads to $g'=0$  but not necessarily to trivial $U_I$
whereas the second condition leads to $f'=0$. 
For the whole axis problem both conditions allow to map physical wave functions
to physical wave functions. For the radial problem 
the bound state wave function $r\cdot \psi(r)$ 
should vanish at the origin like $r^{l+1}$ and decrease at infinity in order
to be regular and normalizable. 
Regularity and normalizability should be preserved in the mapping between
physical wave functions: while $g(r)$ can be assumed to be regular at the
origin, this implies for the function $f(r)$ at 
the origin: 
$$f(r) \sim - \frac{l + 1}{r} \qquad  or \qquad f(r) \sim \frac{l}{r}.$$ 
In the factorization approach 
$W(r) = -\Psi^{(0)\prime}_E / 
\Psi^{(0)}_E\, ,$ where $\Psi^{(0)}_E$ is a solution
(physical or unphysical) of the Schroedinger equation for 
$E=\epsilon_R+i\epsilon_I :$ its behaviour $r^{l+1}$
(physical) or $r^{-l}$ (unphysical) corresponds to the two 
behaviours of $f(r),$ respectively. This expression for the 
superpotential $W(r)$ displays its parametric dependence on
$E=\epsilon_R+i\epsilon_I.$
The first behaviour maps the centrifugal barrier for $l$ to the partner
barrier $l+1$ while the second leads  \footnote{As mentioned in
\cite{acdi} there is an alternative interpretation which considers
still the angular momentum to remain $l$ but allows for the existence of
a repulsive centrifugal like singular potential of dynamical origin.}
 to $l-1$.
So for the radial
case the condition $f = const$ is not compatible with the physical
behaviour at the origin.

In contradistinction, on the line the boundary conditions 
for the wave functions are a less severe constraint
because one can have arbitrary smooth behaviour at the origin and suitable
decrease at $\pm \infty .$

In Sect.2 we derive, from the basic formalism for the construction of 
a SUSY partnership for complex potentials given in (\ref{eq1}) -- 
(\ref{eq4}), the
conditions which allow for a partnership between a hermitean
Hamiltonian and a non-hermitean one. The discussion is formulated 
for the half line (radial problem). 

Section 3 concerns with 
the full line problem (one-dimensional Quantum Mechanics).
We construct complex 
transparent potentials (Subsect.3.1) and give examples of non-polynomial 
"quasi-complex" Hamiltonians 
with real spectrum (Subsect.3.2) including remarks about their symmetry
properties . 

An outlook on the general 
isospectrality between 
complex potentials is given
in Sect.4 with contact to the Zakharov-Shabat problem, 
to scattering problems and to Higher order derivative SUSY transformations.
 
\section*{\large\bf 2.\quad $U^{(0)}_I(r) = 0$ and "quasi-complex"
potentials for the radial problem.}
\hspace*{3ex}

We study a special case of partner relationship in which one partner is real
and the other is complex. The non obvious aspect of this relation 
lies in the fact 
that the spectrum of the complex potential will be real only 
if we are able to satisfy the physical conditions at the origin discussed
above. Otherwise the operator intertwining relations (\ref{intertw}), 
(\ref{intertw2}) involve unphysical solutions of the Schroedinger equation.

We try to find cases where the 
real potential $U^{(0)}_R(r)$ has many bound states 
because it is not astonishing that in the 
scattering regime there is a real continuum spectrum for complex potentials 
as one can verify in the case of a constant complex
potential (cf. also eq.(39) of \cite{baye} and its subsequent complex 
extension). Note incidentally that, at least restricting to real potentials,
the constant potential and the example of \cite{baye} are well known 
standard SUSY partners. 

Eq.(\ref{eq2}) for $U^{(0)}_I(r) = 0$ allows to express 
the function $f(r)$ in
terms of $g(r) :$
\be
f(r) = \frac{g'(r) - \epsilon_I}{2g(r)}; \qquad g(r) \neq 0 , \label{elim}
\ee
which implies:
\ba
U^{(0)}_R(r) &=& \frac{3}{4}\frac{g^{\prime 2}}{g^2} - 
\frac{g^{\prime\prime}}{2g} - \frac{\epsilon_I\cdot g^{\prime}}{g^2} +
\frac{\epsilon_I^2}{4g^2} - g^2 +\epsilon_R ; \label{eqq1}\\
U^{(1)}_R(r) &=& -\frac{1}{4}\frac{g^{\prime 2}}{g^2} + 
\frac{g^{\prime\prime}}{2g} +
\frac{\epsilon_I^2}{4g^2} - g^2 +\epsilon_R ; \label{eqq2} \\
U^{(1)}_I(r) &=& 2g' .  \label{eqq3}
\ea
Thus we can obtain a complex potential $V^{(1)}$ which is equivalent,
in the sense of intertwining relations, to a purely real potential. From 
now on we will call such complex potentials "quasi-complex" potentials.
The formal definition is given in terms of the  equations above.

The requirement that flux is absorbed and not generated leads to the 
dissipative condition $U^{(1)}_I(r) \leq 0 $ i.e. $g'(r) \leq 0 ,$
which is quite common in the context of optical potential studies. 
In this case the previous equations cannot
be solved unless $U^{(0)}_R(r)$ has no bound states. 
This is so because the eigenvalues of the Hamiltonian 
for a complex dissipative potential have negative imaginary parts \cite{baye}
\footnote{The terms not appearing in eq.(8) of \cite{baye} vanish if one
imposes physical boundary conditions for the wave functions.}
\be
Im E_n = \frac
{\bigl [ 2\int_0^{\infty} g^{\prime}(r) |\Psi_n (r)|^2 dr - \frac{i}{2}
\bigl (\bar\Psi_n^{\prime}(r) \Psi_n(r) - 
\bar\Psi_n(r)\Psi^{\prime}_n(r)\bigr )^{\infty}_0 \bigr ]}
{\bigl (\int_0^{\infty} |\Psi_n (r)|^2 dr\bigr )}
, \label{imaginary}
\ee
while the partner (hermitean) Hamiltonian should have the real spectrum.

It is instructive to check how this is realized 
for a class of models for which the real part 
$f(r)$ of the superpotential  coincides with the class of solvable 
models in standard SUSY 
Quantum Mechanics \cite{dabro}:
$$
f(r) = A\cdot tanh\, \alpha r - B\cdot coth\, \alpha r .$$ 
One finds that the condition of dissipativity 
$( g'(r) \leq 0 )$ implies $g'(0)\cdot g'(\infty) > 0$ and
leads to the requirement 
$ B < 0,$ in conflict with the condition 
$ A > B \geq 0$ required for
the existence of bound states for the potential $U^{(0)}(r).$

Not imposing the dissipative condition we provide a variety of examples with
a pure discrete spectra. The centrifugal behaviour of $f(r)$ at the origin 
implies (\ref{elim})  for $\epsilon_I \neq 0$:
\ba
f(r) \sim - \frac{l_0 + 1}{r} \qquad \Rightarrow \qquad
g(r) \sim a\cdot r;\qquad a = \frac{\epsilon_I}{2l_0+3};  \nonumber\\
f(r) \sim \frac{l_0}{r} \qquad \Rightarrow \qquad
g(r) \sim a\cdot r ; \qquad a = - \frac{\epsilon_I}{2l_0-1}. \nonumber
\ea
or for $\epsilon_I = 0:$
$$f(r) \sim \frac{l_0}{r} \qquad \Rightarrow \qquad 
g(r) \sim \mu r^{2l_0}.$$

Given this behaviour at the origin 
for $\epsilon_I \neq 0,$ it is natural to solve 
in terms of the ansatz:
$$g(r) \equiv a\cdot r\phi(r); \qquad \phi(0) = 1;\qquad \phi(\infty) = 0,
$$
while the first condition 
for $\phi (r)$ is obvious, the second one is suggested by the
requirement of having a purely discrete spectrum (\ref{eqq1}),(\ref{eqq2}).

First we construct potentials which have the behaviour
of generalized oscillators at infinity with the mapping which
preserves normalizability and therefore corresponds to strict 
isospectrality.
We can choose among a variety of cases 
\be
\phi(r) = (r^{2m} + 1)^{-1}; \qquad  m \geq 1, \label{example1}
\ee
Asymptotically one obtains for large $r$:
\be
f(r) \sim \frac{\epsilon_I}{2a}\cdot r^{2m-1}; \qquad
U^{(0)}_R(r) \sim f^2; \qquad \Psi_n^{(0)}(r) \sim exp(-\rho r^{2m});
\label{oscill}
\ee
with $\rho$ a real parameter.

Alternatively we can choose:
$$\phi(r) = exp(-\alpha r); \qquad \alpha > 0.$$
Asymptotically for large values of $r$:
$$
f(r) \sim \frac{\epsilon_I}{2ar}\cdot exp (\alpha r); \qquad
U^{(0)}_R(r) \sim f^2(r); \qquad \Psi_n^{(0)}(r) 
\sim exp \bigl(-\rho \int_0^r \frac{exp (\alpha r)}{r}dr\bigr);
$$
Despite the exponential growth of $f(r)$ one can check that
the mapping preserves normalizability because of the extremely
fast decrease of the wave functions.

\section*{\large\bf 3.\quad $U^{(0)}_I(x) = 0 :$ transparent 
and "quasi-complex" potentials on the line.}
\hspace*{1ex}

In Sect.2 we have discussed intertwining relations between real and
complex potentials in SUSY Quantum Mechanics which allowed us
to introduce for the half line, i.e. the radial problem, the concept 
of quasi-complex potentials.
Due to the less stringent conditions at the origin we will be able on
the line to introduce within the class of quasi-complex potentials
 a sub-class of complex transparent potentials, equivalent in the
sense of the intertwining relations to a vanishing potential. 

\subsection*{\normalsize 3.1.\quad Complex transparent potentials on
the line.}
\hspace*{1ex}
We begin from a specific attempt,
the so called transparency problem (cf. for example 2.5 of \cite{soliton}),
to intertwine a complex Hamiltonian $H^{(1)}$
to a free Hamiltonian $H^{(0)}$ with $U^{(0)}_R = U^{(0)}_I = 0.$ We
insert
the condition $\epsilon_R = \epsilon_I = 0$ in eq. (\ref{eqq1}) and
impose its vanishing:
\be
U^{(0)}_R(x) = \frac{3}{4}\frac{g^{\prime 2}}{g^2} -
\frac{g^{\prime\prime}}{2g} - g^2 = 0. \label{transp}
\ee
Making the substitution $g(x) = \pm p^{-2}(x)$ one obtains
$$
\bigl ( p^{\prime 2}(x) + p^{-2}(x) \bigr)^{\prime} = 0
$$
which can be integrated and yields:
\be
g(x) = \frac{a}{1 + a^2(x + b)^2} 
= \mbox{\rm Im}\frac{-1}{x+b+i/a}
\label{aaa}
\ee
in terms of the arbitrary real constants $a \neq 0$ and $b.$
By inserting (\ref{aaa}) in (\ref{eqq2}) we are thus able to
write the "transparent" complex potential
\be
V^{(1)}(x) = 2a^2 \frac{\bigl [ -1 + a^2 (x + b)^2 \bigr ]}
{\bigl [ 1 + a^2(x + b)^2 \bigr ]^2} -
i \frac{4a^3 (x + b)}{\bigl [ 1 + a^2(x + b)^2 \bigr ]^2} =
\frac{2}{(x + b + \frac{i}{a})^2} .
\label{Vtransp}
\ee
This potential gives a trivial $S$-matrix but generates a zero-energy
bound state,
\be
\Psi^{(1)}_0(x) = \frac{C}{x + b + \frac{i}{a}}. \label{Zen}
\ee

The potential $V^{(1)}(x)$ is invariant under PT reflection where the P-parity
inversion
is performed with respect to $x = - b$ and the T-transformation leads to the
complex conjugation of the potential ( see discussion on the role
of PT-invariance in next Subsection, after (\ref{an}) ), the imaginary part 
of the potential is odd under P-inversion.
 
The transparent potential (\ref{Vtransp}) is  a limiting 
case $(\epsilon_R \to
0^-)$ of
the analytic solution of (\ref{eqq1}) with $\epsilon_R < 0$ and still
$\epsilon_I = 0.$
Proceeding as before we can integrate
$$
\bigl ( p^{\prime 2}(x) + p^{-2}(x) + \epsilon_R p^2(x)
\bigr)^{\prime} = 0
$$
and obtain (with $a,b\,$ real constants)
\be
g(x) = \pm p^{-2}(x) = \pm\frac{2\epsilon_R}{a - \sqrt{a^2 - 4\epsilon_R}
\,cosh \bigl [2\sqrt{-\epsilon_R} (x + b) \bigr ]} ,
\label{g}
\ee
where $g(x)$ has in fact no singularity for real $x$ because $\epsilon_R$
is negative.
In the limit $\epsilon_R \to 0^-$ one can indeed check that (\ref{g})
reduces to (\ref{aaa}).
Without giving the details of the derivation we write the resulting complex
transparent potential:
\ba
U^{(1)}_R(x) = 4\epsilon_R \frac{(a^2-4\epsilon_R) -
a\sqrt{(a^2-4\epsilon_R)}
\,cosh(2\sqrt{-\epsilon_R} (x + b))}{[ a -
\sqrt{(a^2-4\epsilon_R)}\,
cosh(2\sqrt{-\epsilon_R} (x + b)) ]^2};\nonumber \\
U^{(1)}_I(x) = 8\epsilon_R \frac{ \sqrt{-\epsilon_R}
\sqrt{(a^2-4\epsilon_R)}\, sinh(2\sqrt{-\epsilon_R} (x + b))}
{[ a -
\sqrt{(a^2-4\epsilon_R)}\,
cosh(2\sqrt{-\epsilon_R} (x + b)) ]^2} .\label{new}
\ea

The total superpotential $W(x)$ can be reexpressed as:
\be
W(x) = -\sqrt{-\epsilon_R} \cdot tanh(\sqrt{-\epsilon_R}(x+b) + i\rho),
\label{Super} 
\ee
if instead of $a$ we introduce a real constant $\rho = 
\pm 1/2 \cdot Arctan(2\sqrt{-\epsilon_R}/ a)$ 
with  $\rho \neq \frac{\pi}{2} (2n+1)$ in order to avoid potentials 
with strong singularities for real values of $x.$ Correspondingly,
(\ref{new}) can be simplified:
\be
V^{(1)}(x) = U^{(1)}_R(x) + iU^{(1)}_I(x) = 
\frac{2\epsilon_R}{cosh^2\bigl(\sqrt{-\epsilon_R}(x+b)+i\rho\bigr)}.
\label{New}
\ee
One can easily check that it vanishes at infinity and has a bound state 
with energy 
$\epsilon_R$ given by
\be
\Psi^{(1)}_{\epsilon_R}(x) =
\frac{C}{cosh\bigl(\sqrt{-\epsilon_R}(x+b)+i\rho\bigr)}.
\ee
The potential is invariant under PT transformation and its imaginary
part is P-odd.

It is worth to note that there is here a very specific reason 
why the complex potential can be obtained from well-known
real transparent potential $cosh^{-2}$ by a complex shift of the 
coordinate $x.$ Indeed, the Hamiltonian $H^{(0)}$
with constant potential is invariant under an arbitrary (even
complex) translation in $x.$ Factorizing $H^{(0)},$ we can thus obtain
a variety of superpotentials which differ by this arbitrary shift of $x.$
Correspondingly, we have a variety of  superpartner Hamiltonians $H^{(1)}$
with potentials of the form (\ref{New}) both with the real and complex shifts
of $x$. In the framework of SUSY 
transformations, Hamiltonians of this variety can be connected each to
other by the second order SUSY transformations \cite{acdi} via the intermediate
Hamiltonian $H^{(0)}.$ 

It is necessary to stress
that in general a complex shift of $x$ can lead to drastic and non trivial
consequences. In particular, it can change the analytical 
properties of the potential (and of the wave functions): compare
(for real $x$) the nonsingular 
potential (\ref{Vtransp}) with its very singular  real analogue
$2/(x+b)^2.$ On the contrary, SUSY transformations with
nonsingular superpotentials represent the tool to build
complex potentials with spectral properties under 
control (mapping eigenfunctions into eigenfunctions).  

Transparent potential for $\epsilon_R > 0$ and $\epsilon_I = 0$
can be formally deduced from (\ref{g}) with hyperbolic cosine appropriately
replaced by trigonometric cosine. Again there are no singularities
in $g(x).$ The
resulting potential has oscillatory behaviour, apparently compatible
with transparency, taking the limiting case $\epsilon_R \to 0^+$
leads to (\ref{aaa}).

\subsection*{\normalsize 3.2.\quad Non-polynomial quasi-complex potentials
on the line.}
\hspace*{1ex}

In this Subsection we construct quasi-complex potentials on the line
restricting ourselves to potentials bounded from below at infinity
\footnote{See \cite{afterLamb} for discussion of eigenvalue problem
for real potentials unbounded below at infinity.}.
Let us first illustrate why we are unable to construct 
quasi-complex polynomial
Hamiltonians by the intertwining relations (\ref{intertw}).
If we assume that $g(x)\sim x^n$
and we require the potentials to be bounded from below,
from Eqs.(\ref{eq1}) --
(\ref{eq4}) we can derive that $f(x)$ grows
at $\infty$ as a power or faster.
This excludes the possibility to obtain $U^{(0)}_I(x)=0$ with this
ansatz, at least with the supercharges (\ref{supercharge}),
(\ref{supercharge2}), since eq.(\ref{elim}) is not compatible with
the growth of $f(x)$  at $\infty .$
For completeness we mention that the cases of constant $f(x)$ or $g(x)$
are presented in Sect.4 and that some remarks concerning non-hermitean
polynomial Hamiltonians will be given at the end of this Subsection.

As a first class of examples of non-polynomial quasi-complex potentials
it is possible
to extend the generalized oscillator
(\ref{example1}), (\ref{oscill}) to the whole axis.
As for the radial case one obtains
strict isospectrality since the mapping preserves normalizability.

For real potentials it is well known \cite{dabro} that solvable models
can be generated by assuming a superpotential $W(x)\sim tanh \, \alpha x.$
Here we investigate  the ansatz for $g(x)$ to be inserted in
(\ref{eqq1}),(\ref{eqq2}),(\ref{eqq3}):
\be
g(x) = \beta\cdot tanh(\alpha x).  \label{th}
\ee
This leads naturally to a definite sign for $g^{\prime}(x),\,$cf.(\ref{eqq3}),
 for example dissipativity if $\alpha \beta < 0.$
>From eq.(\ref{elim}) one can derive that $f(x)$ at $\pm\infty$
becomes constant.

Imposing furthermore the absence of centrifugal like singularities at
the origin,
 implies a numerical condition
on $\epsilon_I,$ namely $\epsilon_I = \alpha\cdot\beta.$

 From general arguments
dealing with a dissipative case (\ref{imaginary}) strict isospectrality
implies that there cannot exist bound states if the mapping preserves
normalizability. Indeed from isospectrality, one would deduce that in the
same time
eigenvalues have imaginary part for one potential and are real for the other,
but they should also be equal. Of course all our arguments are to be intended
to be true in general apart from the possible existence of zero modes.

First we argue that no bound state exists for
$E=\epsilon_R+i\epsilon_I.$
This can be derived by taking into account that the
zero modes of the operator $q^-$ (\ref{supercharge2}) correspond to the
solutions of the Schroedinger equation for complex spectral parameter
$E=\epsilon_R+i\epsilon_I.$ From this one can conclude that the
asymptotic behaviour of $f(x)$ is not compatible with the existence
of a normalizable solution for that energy.

The absence of bound states at other energies is however now
not necessarily paradoxical since the potential is finite.

Although the previous arguments provide a proof that there cannot
be bound states it is useful to develop an independent proof
of the absence of bound states for the real potential
\be
 U^{(0)}_R(x) = \frac{3}{4}\frac{g^{\prime 2}}{g^2} -
\frac{g^{\prime\prime}}{2g} - \frac{\epsilon_I g^{\prime}}{g^2} +
\frac{\epsilon_I^2}{4g^2} - g^2 +\epsilon_R . \label{real}\\
\ee

The task of a proof of absence of bound states is normally not an easy
one for a  generic potential \cite{simon}.
 In our case, however, performing the substitution
of (\ref{th}) in
(\ref{real}) one can directly conclude that the potential is strictly
repulsive (independently of the value of $\alpha$ and $\beta$)
and proportional to $sech^2\alpha x,$ apart from an energy shift.
A discussion of this potential in the radial case can be found in
\cite{baye}.
In contradistiction $U^{(1)}_R(x)$
can be attractive. In particular one can choose the constants
$\alpha = -2\beta /\sqrt{3}$ such as to make $U^{(1)}_R(x)=$const,
which renormalizes the real part of the energy.
This last choice leads to an interesting relation between a genuine
repulsive real potential and an absorptive purely imaginary potential.

In this Section and in Sect.2 we have constructed quasi-complex
potentials which have nontrivial
imaginary part but the spectra of the corresponding Hamiltonians
are real apart from (possibly) one level. From eq.(\ref{imaginary})
it is clear that in some peculiar average sense the imaginary
part $U^{(1)}_I(x)$ of the potential vanishes:
we stress that Eq.(\ref{imaginary}) vanishes for all eigenfunctions,
thereby reflecting a global property of the potential \footnote{It is useful
to remark
that introducing explicitly $\hbar$ in the Schroedinger equation and
taking the classical limit $\hbar \to 0$ the Eq.(\ref{eqq3}) vanishes
and (\ref{eqq1}) coincides with (\ref{eqq2}) because of the last
three terms.}.

We give a more technical illustration of such a type
of global property and discuss how it can be implemented on the full line
by a suitable Gaussian ansatz with $\epsilon_I = 0:$
$g(x) = \gamma\cdot exp(-\alpha x^{2n})$ with positive $\alpha$
and $\gamma,$ qualitatively to be interpreted in terms of "range"
and "strength" parameters.
Correspondingly $f(x) = g'(x)/2g(x) = - \alpha n x^{2n-1}.$
By this ansatz  we
construct the real potential $V^{(0)}(x)$ and the complex potential
$V^{(1)}(x)$ according to (\ref{eqq1}) -- (\ref{eqq3}).

The potentials one obtains are:
\ba
V^{(0)}(x) &=& U^{(0)}_R(x) = \alpha^2n^2x^{4n-2} + \alpha n(2n-1)x^{2n-2}
- \gamma^2\cdot exp(-2\alpha x^{2n}) + \epsilon_R; \nonumber\\
V^{(1)}(x) &=& U^{(1)}_R(x) + i U^{(1)}_I(x) =
\alpha^2n^2x^{4n-2} - \alpha n(2n-1)x^{2n-2} -\nonumber\\
&-& \gamma^2\cdot exp(-2\alpha x^{2n}) + \epsilon_R -
 4i\alpha\gamma nx^{2n-1} exp(-\alpha x^{2n}).
\label{an}
\ea
Both potentials are regular and growing at infinity leading to a discrete
real spectrum since the supercharges respect in this case the
normalizability of the wave functions.

We make explicitly use of the supersymmetry to connect, apart from
a normalization factor, the eigenfunctions $\Psi^{(0)}(x)$ and
$\Psi^{(1)}(x)$ and rewrite (\ref{imaginary})
\be
Im E_n = \frac{\int_{-\infty}^{\infty}2 g^{\prime}(x)
|\Psi^{(1)}_n (x)|^2 dx}
{\bigl (\int_{-\infty}^{\infty} |\Psi^{(1)}_n (x)|^2 dx\bigr )}
\label{imaginary1}
\ee
with
\be
\Psi^{(1)}(x) \sim (\partial + f(x) + ig(x)) \Psi^{(0)}(x).
\label{psi}
\ee
With our choices
$Im E_n = 0$ because $|\Psi^{(1)}_n (x)|^2$ is even
under parity while $g'(x)$ is odd.
Indeed $H^{(0)}$ is invariant separately for parity P and time
reversal (complex conjugation) T inversions while $H^{(1)}$ is
invariant only under the combined PT transformation.
So $\Psi^{(0)}(x)$ has definite parity and can be chosen real
whereas $\Psi^{(1)}(x)$ is complex and has no definite parity, but
by the previous expression (\ref{psi}) we control the parity invariance
of $|\Psi^{(1)}_n (x)|^2.$ Similar arguments hold also for
 the case of transparent complex potentials (\ref{Vtransp}),
(\ref{New}). 

Heuristically one can speculate that in some suitable limit 
of $\alpha$ small
and $\gamma$ large the purely polynomial terms can 
be neglected. Restricting to a finite interval of $x,$ the
exponentials can be, roughly speaking, approximated by unity.
Apart from an energy downward shift, taking such a limit for 
$\alpha ,\gamma$ in a finite interval of $x,$
the dominating term is $U^{(1)}_I(x)=-4i\alpha\gamma nx^{2n-1}.$

The invariance properties of the model (\ref{an}) are shared by systems
with polynomial potentials
$V = A x^{2m} + iB x^{2n+1}$ on the whole axis.  
The Hamiltonian $H$ is as before not parity invariant and
not time reversal (complex conjugation) invariant but still 
possesses an antiunitary PT invariance. 
This property implies
the reality of the matrix elements of $H^{(1)}$ in a suitable basis 
(for a constructive proof
cf. \cite{bohr}). This obviously is consistent 
with the vanishing of (\ref{imaginary1}).    

It is remarkable that this type of systems was discussed recently
in a framework of a supersymmetric quantum field theory \cite{bender}
(in connection with the studies of the so-called Lee-Yang zeros)
and the spectrum was found numerically (Blencowe et al. \cite{bender}) 
to be real and bounded below.

\section*{\large\bf 4.\quad Outlook and perspectives.}
\hspace*{1ex}

In the previous Sections we have illustrated the consequences
of the intertwining relations by discussing several examples
in considerable detail.
In this last Section we want to mention briefly some topics
which can also be treated in the formalism we have discussed above.
For conciseness we will not illustrate in detail various possible
applications but only outline few paths which can be explored.

\subsection*{\normalsize 4.1.\quad Scattering.}
\hspace*{1ex}
 
Considering potentials with continuum spectrum (see for
example \cite{sitenko}),
we start from asymptotic states of a real potential $U^{(0)}_I(r)=0.$
These scattering states of the real potential $U^{(0)}_R(r)$
in partial wave $l_0,$ asymptotically at $\infty$ 
(for the notations cf. \cite{acdi}) read:
\be
\Psi^{(0)}(r) \sim exp(-ikr) + (-1)^{l_0+1} S_{l_0}^{(0)}(k) exp(ikr) .
\label{scat}
\ee
We illustrate for a specific value 
of $k=k_0,$ how one can generate for the complex
partner Hamiltonian a partner wave function $\Psi^{(1)}(r)$ with 
no ingoing wave , a condition commonly
related to bound or quasi-stationary states (resonances) in the case of
real potentials. We choose $g(\infty) = k_0 + \alpha\cdot exp (-\beta r).$
Assuming $\epsilon_I = 0,$ from (\ref{elim}) we obtain $f(\infty) = 
-(\alpha\beta /2k_0)\cdot exp(-\beta r).$ By
applying $q^-(\infty) = (\partial +f(\infty) + ig(\infty))$ to 
$\Psi^{(0)}(r),$ the ingoing wave in $\Psi^{(1)}(r)$ is exponentially
damped. The energy associated to (\ref{scat})
for $k=k_0$ is indeed $E(k_0) = k_0^2 + U^{(0)}_R(\infty) = \epsilon_R,$
as expected from standard SUSY Quantum Mechanics, cf. Andrianov et al. in 
\cite{review}.  
A similar treatment for $k = - k_0$ leads to an exponential damping 
of the outgoing wave in  $\Psi^{(1)}(r)$ which can be interpreted
as black sphere or strong absorption limit for this specific value of $k.$  

Even
if in the present paper the main attention has been given to the case
of strict isospectrality which respects normalizability, there are 
other interesting cases where intertwining relations do not respect
normalizability because of the behaviour of $f$ and $g.$ 
For a suitable falloff of the complex potential, bound and quasi-stationary 
wave functions are for large values of $r,$ correspondingly, 
the decreasing and increasing solutions of the (asymptotic) equation
$-\Psi^{\prime\prime} = (E_R + iE_I) \Psi .$  
One could
consider, for example, the SUSY partnership between a complex potential,
leading to a bound state with positive real energy and negative imaginary
part of energy \cite{baye}, and another complex potential for which
at this complex energy one has a non-normalizable quasi-stationary 
(resonant) wave
function \cite{resonance}. This can be achieved with the same 
techniques outlined above which produce 
an exponentially decreasing state by application of a suitable
supercharge $q$ to a non-normalizable state (creation of bound state). 
Similar results can be 
obtained by complex rotation approach (for example, \cite{sofianos}
and references quoted therein). Alternatively, the reverse mapping from
normalizable to non-normalizable (cancellation of bound state) 
can be realized because (\ref{elim}) does
not hold for the case of complexity of both potentials.

For scattering on the line one can construct SUSY transformations
for $\epsilon_I=0$
which also lead to an extinction of the ingoing wave. On the other
hand transformations with $\epsilon_I\neq 0$ (an explicit realization
can be obtained for the potential defined by (\ref{th})) formally 
provide the extinction only for complex values of $k.$  

\subsection*{\normalsize 4.2.\quad Isospectrality between complex potentials.}
\hspace*{1ex}

This topic has been thoroughly investigated by Baye et al.\cite{baye}
with special emphasis on the construction of phase equivalent potentials.
In our approach
starting from the equations (\ref{eq1}) -- (\ref{eq4}) one can 
construct families of isospectral complex potentials if $f(y)$ and
$g(y)$ are such as to preserve the normalizability in the mapping
(cf. discussion in Sect.2 and Sect.3 of this paper and the 
approach of \cite{baye}). 
This allows in principle to investigate
the interplay between repulsion $f^2(y)$ (cf. eq.(\ref{kappa}) below) 
in the real part of the
potential and the absorptive potential (depending on $f(y)$ 
both for bound states and 
the scattering problem). 

For simplicity we consider the case of the whole
axis which e.g. is appropriate for physical applications concerning 
solitons (cf. the so-called Zakharov - Shabat problem and its
generalizations in 3.9 of \cite{soliton}). 
The most simple example has already been 
mentioned in Sect.1 and can be formulated as follows:
\ba
g(x)\equiv \kappa = real\quad const;\nonumber\\ U^{(0)}_I(x)=U^{(1)}_I(x)=
2\kappa f(x) +\epsilon_I;\nonumber \\
U^{(0),(1)}_R(x)=\mp f'(x)+f^2(x)+\epsilon_R-\kappa^2.
\label{kappa}
\ea 
As a particular case we can solve the above equation for $U^{(0)}_R(x)=0$
or $U^{(1)}_R(x)=0,$ this is the well known analytically solvable
case of the Riccati equation \cite{ince}.

We can also discuss a mapping\footnote{In Sect.2 
we have explained why this cannot be done on the half axis.}
which does not affect the real part of the potential with 
$f(x)\equiv \lambda =$ real const: 
\ba
U^{(0)}_R(x)=U^{(1)}_R(x)=\lambda^2-
g^2(x) +\epsilon_R; \nonumber \\
U^{(0),(1)}_I(x)=\mp g'(x)+2\lambda g(x)+\epsilon_I.
\label{lambda}
\ea 
One can notice that for any $g(x)\,$, apart from an energy shift,
$U^{(0)}_R(x)=U^{(1)}_R(x)\,$ are attractive. 

Ansaetze leading to rather simple expressions for eq.(\ref{lambda})
are $ g(x) \sim sech\,x$ or 
$ g(x) \sim [ 2\,cosh\,x \cdot sech\,2x - 1].$

Applications involving
$\lambda = 0$ given in \cite{neutrino} include problems
with two-state systems and some phenomenological models for quark-gluon 
plasma physics and neutrino propagation in matter.
One can also notice that the Zakharov-Shabat problem, cf. 2.12 of 
\cite{soliton}:$$n_1^{\prime}(x) +
iu(x) n_1(x) = \zeta n_2(x); \quad n_2^{\prime}(x) -
iu(x) n_2(x) = -\zeta n_1(x)$$ reduces to the Schroedinger
equation with $V^{(0)}(x)$ of eq.(\ref{lambda}) for $\lambda = 0$ :
$$-n_1^{\prime\prime} - (u^2 + iu^{\prime})n_1 = \zeta^2n_1. $$
In this respect we would like to remark that in the 
mathematical literature there are interesting problems \cite{soliton} 
(Sect.1.5
and Sect.5.1) where the imaginary part $U_I(x)$ of complex potential is not 
everywhere dissipative. 

As a typical example, $$U_I(x) \sim - 
\frac{e^{-x}}{cosh\, 2x} (1 + 2\, tanh\, 2x)$$ results from
 a function $g(x) \sim exp(-x)\cdot sech\, 2x$ with $f(x)=0.$
In this case $U_I^{(1)}(x)$ and $U_I^{(0)}(x)$ have opposite sign apart
possibly for a constant $\epsilon_I$ and instead 
 $U_R^{(1)}(x)$ and $U_R^{(0)}(x)$ are equal.  

As another exercise it is possible to construct the potentials 
if $f(x)$ in (\ref{kappa})
and $g(x)$ in (\ref{lambda}) are assumed to be of Gaussian shape, 
like in Subsect. 3.2.

\subsection*{\normalsize 4.3.\quad Higher order derivative 
supercharges.}
\hspace*{1ex}

By suitably iterating intertwining relations with different $f$ and $g$
which strictly preserve isospectrality, as explained above, one can
generate an infinite class of quasi-complex potentials.  Finally, one can
also find an intertwining relations where the Hamiltonian $H^{(n)}$ is
again hermitean: this provides an explicit construction
of a Higher order irreducible SUSY transformations as discussed in
\cite{acdi}, i.e. transformations which cannot be represented as a sequence
of two standard (with real superpotentials) first order transformations. 
The concept of irreducibility depends on the
class of intermediate Hamiltonians which are allowed. What was an
irreducible transformation allowing only hermitean Hamiltonians
\cite{acdi} can become a transformation which effectively is reducible
allowing non-hermitean intermediate Hamiltonians. In a similar way one
can discuss second order complex SUSY transformations
between two non-hermitean dissipative Hamiltonians. Again the intermediate
Hamiltonian is not necessarily dissipative everywhere.

\section*{\normalsize\bf Acknowledgements}

This work was made possible by support provided by Grant of
Russian Foundation of
Basic Researches (No.96-01-00535) and GRACENAS Grant (No.95-0-6.4-49).
The collaboration has been
also financed by the agreement of S.Petersburg State University, IHEP, PNPI
and INFN. Support from University Paris 7 Denis Diderot and from 
CIMO (Finland) is
gratefully acknowledged by one of us (M.V. I.). J.-P. D. benefitted
from illuminating discussions with Dr. B. Grammaticos.
The authors are grateful to referee for stimulating remarks concerning 
Subsection 3.1.

\end{document}